\documentclass[aps,preprint,showpacs, showkeys]{revtex4}%
\usepackage{amsfonts}
\usepackage{amsmath}
\usepackage{amssymb}
\usepackage{graphicx}%
\providecommand{\U}[1]{\protect \rule{.1in}{.1in}}

\begin{document}
\title[ ]{On the $i\phi^{3}$ $\mathcal{PT}$-symmetric Scalar Field Theory}
\author{${}^{1,2}$ Abouzeid M. Shalaby }
\email{amshalab@mans.edu.eg}
\affiliation{${}^{1}$Physics Department, Faculty of Science, Mansoura University, Egypt
\\${}^{2}$ Physics Department, Faculty of Science, Qassim University, Saudi Arabia}
\keywords{non-Hermitian models, $\mathcal{PT}$-Symmetric theories, Conformal field theory,Solitons.}
\pacs{12.90.+b, 12.60.Cn, 11.30.Er}

\begin{abstract}
In this work, we show that, for the $i\phi^{3}$ scalar field theory, their
exists a contradiction between the assumption that the field is real and the
fact that the quantized as well as the classical fields have to satisfy the
Klein-Gordon equation. In solving the Klein-Gordon equation for the theory
under investigation, we realized that the field is a pure imaginary solitary
wave which spoils out the non-Hermiticity of the theory. Thus, instead of
being non-Hermitian, the $i\phi^{3}$ scalar field theory is a kind of a
Hermitian-Lee-Wick theory which suffers from the existence of the famous ghost
states and instability problems. We applied a Canonical transformation to
obtain a Non-Hermitian and non-$\mathcal{PT}$-symmetric representation which
leads to the invalidity of the previous trials in the literature to cure the
ghost states problem. Moreover, the solitonic solution is a non-topological
one which is a very strange result to appear for a one component field theory.
To account for this strange result, we conjecture that the $i\phi^{3}$ scalar
field theory has an equivalent Hermitian and non-Lee wick theory that have a
conserved Noether current.

\end{abstract}
\maketitle

The existence of challenging problems in physics makes the investigation of
new trends in physics more than important. Among the new ideas that deserve
the draw of our attention is the emergence of possible physical applications
of some non-Hermitian models
\cite{tes,zad1,bend1,Conj,bendg1,bend2,bend3,zadah,cop,abophi41p1,spect,spect1,aboeff,abomet}%
. Some of these models show up interesting properties like asymptotic freedom
in the $\mathcal{PT}$-symmetric $-\phi^{4}$ scalar field theory
\cite{Symanzik, bendf, Frieder}. This property by itself strongly recommends
the employment of the $\mathcal{PT}$-symmetric $-\phi^{4}$ theory to play the
role of a strongly interacting Higgs mechanism in the standard model of
particle interactions \cite{abohir}. However, unlike the quantum mechanical
non-Hermitian models, the progress in the study of non-Hermitian field
theoretic models is slow due to the existence of two main technical problems. The
first problem is that the complex contour method applied successfully in
quantum mechanical problems \cite{bend1f,bend2f,bend3f,bend4,bend5,jones} is
hard (if not impossible) to be applied for the quantum field models. The
second problem is the lack of existence of a simple algorithm for the
calculation of the metric operator for field theories \cite{bend6}. In fact,
for the calculation of physical observables in a non-Hermitian theory (with
real spectrum), the metric operator calculation is indispensable. Concerning
the first problem, we have shown that the famous effective field approach can
be applied successfully for such theories \cite{aboeff}. For the second
problem, we introduced a new ansatz for the metric operator which is local in
the fields and can be extended to higher orders easily \cite{abomet,abometf} .

A prototype example of a $\mathcal{PT}$-symmetric \footnote{ We will show that this terminology is no longer correct for the theory under investigation.} field theory is the
$i\phi^{3}$ scalar field theory. Against all the previous treatments carried
out to study the model \cite{abomet,bend4,bend6}, in this work, we show that
the model is in fact a Lee-Wick \cite{lee1,lee2} Hermitian theory. The problem
is manifested by the priori assumption of the reality of the field without any
reference to the solution of the corresponding Klein-Gordon equation. In fact,
the quantized as well as the classical fields have to satisfy the Klein-Gordon
equation for the theory under investigation. For the $i\phi^{3}$ scalar field
theory, it is well known that \cite{Greiner},
\begin{equation}
\overset{\cdot}{\pi}=-i\left[  \pi,H\right]  =\left(  \nabla^{2}-m^{2}%
-3ig\phi \right)  \phi,
\end{equation}
where $\pi$ is the canonical momentum field conjugate to $\phi$. In other
words, the quantized field $\phi$ has to satisfy the klein-Gordon equation.
This important realization, will lead to a new vision which is introduced in
this work for the first time about the treatment of the $\mathcal{PT}%
$-symmetric $i\phi^{3}$ scalar field theory. The new vision will prove the
validity of our ansatz in Ref.\cite{abomet} as well as will invalidate the
other trials carried out for the $C$ operator calculations \cite{bend4,bend6}
of the $\mathcal{PT}$-symmetric $i\phi^{3}$ scalar field theory.

The back bone for the calculation of observables in non-Hermitian theories
with real spectra is the metric operator $\eta$. This operator has the
property ~$H^{\dagger}=\eta H\eta^{-1}$, where $H$ is a Hamiltonian operator
and $H^{\dagger}$ is its adjoint \cite{spect,spect1}. Here $\eta$ is
Hermitian, invertible and positive definite metric operator. In fact, $\eta$
is not unique \cite{unique00,unique0,unique1,unique2,unique3,unique4} and thus
one can find different forms for $\eta$ for the same Hamiltonian operator $H$.
While the metric operator calculations go easily in the quantum mechanical
case, the calculations are very complicated for the quantum field versions.
Moreover, the accomplished calculations which so far are believed to be valid
are non-local and can be done in a closed form but for first order
approximation and for $1+1$ space-time dimensions only \cite{bend4,bend6}.

\begin{figure}[ptb]
\begin{center}
\includegraphics{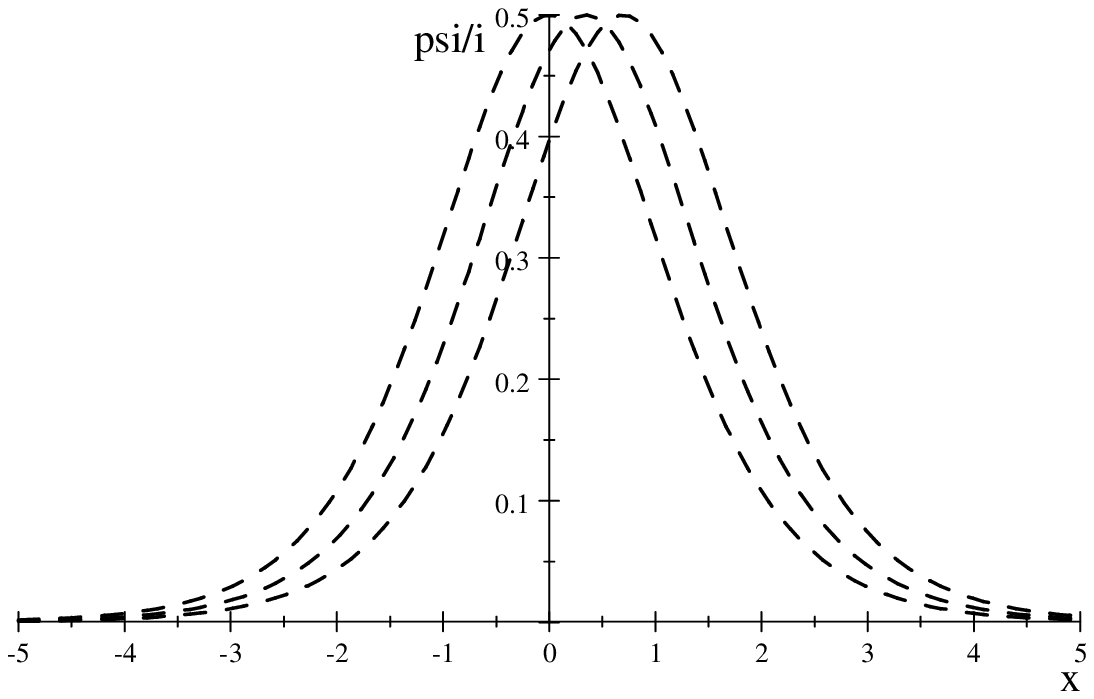}
\end{center}
\caption{The field $-i\psi$ (solitary wave) as a function of $x$ at $t=0$,
$t=0.5$ and $t=1$ for $v=0.7$ from left to right respectively.}%
\label{solit0}%
\end{figure}

The lack of existence of a simple form for metric operator in field theory may
lead to the following question; is the extension from the quantum mechanical
version to the quantum field version can go straight as done in the literature
\cite{bend4,bend6}?. To answer this question, let us consider the quantum
mechanical theory with Hamiltonian $H$ such that;
\[
H=\frac{p^{2}}{2}+\frac{1}{2}m^{2}x^{2}+igx^{3},
\]
where $m$ is the mass parameter and $g$ is the coupling constant. In the
literature, it is assumed that $x$ is real and thus the Hamiltonian is
non-Hermitian but $\mathcal{PT}$-symmetric which means that the spectrum is
real. For the quantum field version of this theory, we have the Hamiltonian
form;
\[
H=H_{0}+gH_{I}%
\]%
\begin{align}
H_{0}  &  =\frac{1}{2}\int d^{3}x\left(  \pi^{2}+\left(  \nabla \phi \right)
^{2}+m^{2}\phi^{2}\right)  ,\nonumber \\
H_{I}  &  =i\int d^{3}x\phi^{3}, \label{ham1}%
\end{align}
where $\phi$ is the field operator and $\pi$ is its canonical conjugate. In
fact, $\phi$ does not represent an observable and one can not have a priory
assumption that it is real as in the case of the quantum mechanical version
where $x$ can be taken real as it represents an observable. In fact, a priori
assumption of the reality of the field $\phi$ in the $\mathcal{PT}$-symmetric
field theory with cubic interaction will lead to a contradiction. To explain
this, we consider the equation%
\begin{equation}
\overset{\cdot}{\pi}=-i\left[  \pi,H\right]  ,
\end{equation}
which is nothing but the klein-Gordon equation for the field $\phi$. In other
words, the quantized field for the $i\phi^{3}$ scalar field theory has to
satisfy the klein-Gordon equation of the form;%
\begin{equation}
\frac{\partial^{2}\phi}{\partial t^{2}}-\nabla^{2}\phi+m^{2}\phi+3ig\phi
^{2}=0.
\end{equation}
In the rest frame and in $1+1$ space-time dimensions, this equation can take
the form;
\begin{equation}
-\frac{d^{2}\phi}{dx^{2}}+m^{2}\phi+3ig\phi^{2}=0.
\end{equation}
in multiplying by $\frac{d\phi}{dx}$, one can have the relation between $\phi
$ and $x$ as
\begin{equation}
x=\int \frac{1}{\sqrt{m^{2}\phi^{2}+2ig\phi^{3}}}d\phi,
\end{equation}
or
\begin{align}
\phi \left(  x\right)   &  =\frac{2im^{2}}{g}\frac{e^{mx}}{\left(
e^{mx}+1\right)  ^{2}},\nonumber \\
&  =\frac{im^{2}}{2g}\operatorname*{sech}{}^{2}\left(  \frac{mx}{2}\right)  .
\end{align}
For the time dependent solution, \ one can boost the static solution above to
get the result;
\begin{align}
\phi \left(  x,t\right)   &  =\frac{2im^{2}}{g}\frac{e^{m\gamma \left(
x-vt\right)  }}{\left(  e^{m\gamma \left(  x-vt\right)  }+1\right)  ^{2}%
},\nonumber \\
&  =\frac{im^{2}}{2g}\operatorname*{sech}{}^{2}\left(  \frac{^{\gamma m\left(
x-vt\right)  }}{2}\right)  . \label{fldt}%
\end{align}
where $\gamma$ is the Lorentz factor and $v$ is the velocity. One can easily
check that the above solution satisfies the Klein-Gordon equation given by;%
\[
\frac{\partial^{2}\phi}{\partial t^{2}}-\frac{\partial^{2}\phi}{\partial
x^{2}}+m^{2}\phi+3ig\phi^{2}=0.
\]
Apart from the operator characteristic of the field which is employed as
Fourier coefficients in the Fourier transform of the field, the important
information in this solution is that the quantized field is a pure imaginary
solitary wave as shown in Fig.\ref{solit0}. This realization is important as
solitary waves are candidates for the description of Hadrons as they retain
their shapes after collision \cite{ram}.

The anti-soliton solution, $\phi^{\dagger}=\frac{-im^{2}}{2g}%
\operatorname*{sech}{}^{2}\left(  \frac{mx}{2}\right)  $, belongs to the
theory with opposite sign of the coupling ($-ig\phi^{3}$). \ With the soliton
\ for $ig\phi^{3}$ theory and the antisoliton solution for $-ig\phi^{3}$
theory, both theories merge into an equivalent one Lee-Wick theory. Since the
soliton solutions $\phi$ and $\phi^{\dagger}$ have the boundary conditions;%
\[
\phi_{\pm \infty}=0,\text{ \  \ }\phi_{\pm \infty}^{\dagger}=0\text{,}%
\]
they are non-topological ones. This is a very strange result because
non-topological solitons are characterizing theories with more than one
component \cite{ram}. Accordingly, one may conjecture that the theory under
investigation has an equivalent Hermitian theory which includes both $\phi$
and $\phi^{\dagger}$ with a conserved Noether current.

In view of the features of the obtained solution in Eq.(\ref{fldt}), the
Hamiltonian in Eq.(\ref{ham1}) is $\mathcal{PT}$-symmetric as well as
Hermitian and thus the metric operator is the Identity operator which means
that all the previous trials to calculate the $C$ operator for the theory
at hand were invalid \cite{bend4,bend6}. However, this Hamiltonian suffers
from the existence of ghost states ( as a Lee-Wick theory) and stability problems. In fact, the
presence of such problems does not mean that the theory is physically
unacceptable. For instance, Bender \textit{et.al} have treated the swanson
model and the Lee model for which such kind of problems exist
\cite{noghost,lee}.

To start the algorithm of curing the ghost states and instability problems in
the $i\phi^{3}$ model, we use the map
\[
\zeta \left(  x\right)  =\int dx\exp \left(  \frac{i\ln b}{2}\left(  \pi \left(
x\right)  \phi \left(  x\right)  +\phi \left(  x\right)  \pi \left(  x\right)
\right)  \right)  ,
\]
where $b$ is a constant $C-number$. In using Baker--Campbell--Hausdorff
formula, one can obtain the following relations;%
\begin{align*}
\zeta \left(  x\right)  \phi \left(  y\right)  \zeta^{-1}\left(  x\right)   &
=b\phi \left(  x\right)  ,\\
\zeta \left(  x\right)  \pi \left(  y\right)  \zeta^{-1}\left(  x\right)   &
=\frac{\pi \left(  x\right)  }{b},
\end{align*}

\bigskip Accordingly,%
\begin{equation}
\zeta \left(  x\right)  H\left(  y\right)  \zeta^{-1}\left(  x\right)
=-\frac{1}{2}\int d^{3}x\left(  \left(  \nabla \phi \right)  ^{2}+\pi^{2}%
+m^{2}\phi \right)  +\int d^{3}x\phi^{3}, \label{ham3}%
\end{equation}
where $b=-i$. One may concludes that this form is equivalent to the form in
the literature with the assumption that the field is real from the very
beginning. In fact, this conclusion is not correct as the Hamiltonian in
Eq.(\ref{ham3}) is no longer $\mathcal{PT}$-symmetric and thus the $C$-operator
regime followed in the literature collapses \cite{bend4,bend6}. As the theory
has a real spectrum, one can find an equivalent Hermitian as well as positive
definite theory via the calculation of the metric operator.

In Refs. \cite{abomet,abometf}, we introduced a new ansatz for the metric
operator for $i\phi^{3}$ scalar field theory and showed that the $Q_1$ operator has the form;

\begin{equation}
Q_{1}=C_{1}\int d^{d}z\pi^{3}(z)+C_{2}\int d^{d}z\phi(z)\pi(z)\phi
(z)+C_{3}\int d^{d}z\nabla \phi(z)\pi(z)\nabla \phi(z),
\end{equation}
where $C_{i}$ are real parameters to be adapted for $Q_{1}$ to satisfy
Eq.(\ref{1st}) below. Note that, the relation $H^{\dagger}=\eta H\eta^{-1}$
has to exist and thus the first order correction for the metric operator can
be obtained from the relation;
\begin{equation}
-2H_{I}=\left[  -Q_{1},H_{0}\right]  , \label{1st}%
\end{equation}
where
\begin{align*}
\eta &  =\exp(-Q),\\
Q  &  =Q_{0}+gQ_{1}+g^{2}Q_{2}++g^{3}Q_{3}+...
\end{align*}

\bigskip Now, back to the soliton solution, in $1+1$ space-time dimensions,
one can obtain its mass as;%
\begin{align*}
M  &  =-\frac{1}{4}\frac{m^{6}}{m}\int_{-\infty}^{\infty}\frac{\cosh^{2}%
\frac{-1}{2}x-1}{g^{2}\cosh^{6}\frac{-1}{2}x}dx\\
&  =-\frac{2}{15}\frac{m^{5}}{g^{2}},
\end{align*}
with the classical energy given by $E=\gamma M$. This means that the soliton
bears a particle characteristics. In fact, this is not the first time to have
negative masses for a $\mathcal{PT}$-symmetric theory \cite{negmass} but its
appearance here in the classical theory is a reminiscent  of the ghost-states but
certainly it would be cured in the quantized one via the use of the metric
operator. Also, this results agrees with the negative central charge of a
closely related model studied in Ref. \cite{muss}.

For Higher space-time dimensions one gets the the solution;%

\[
\phi \left(  x,y,z\right)  =\frac{im^{2}}{2g}sech^{2}\left(  \frac{m\left(
x+y+z\right)  }{2\sqrt{3}}\right)  .
\]
Although this is finite everywhere in the space, the classical energy is
infinite for the higher dimensions which agrees well with Derrick's theorem
\cite{Derk}.

For the quantization of the classical sliton, one can expand the quantized
field around the static solution and use the traditional quantum field methods
to calculate the amplitudes \cite{ram}, which will be a research topic for us
in the near future.

To conclude, We discovered that the now known as $\mathcal{PT}$-symmetric and
non-Hermitian scalar field theories are in fact a kind of Hermitian Lee-Wick
theories. For that, we have solved the Klein-Gordon equation for the
$i\phi^{3}$ scalar field theory. For this this theory, both the classical and
the quantized fields have to verify the Klein-Gordon equation. We found that
the field is pure imaginary and thus spoils out the non-Hermiticity of the
theory. Accordingly, the theory is Hermitian but suffers from existence of
instability and ghost states problems. To treat these problems, we applied a
Canonical transformation to transform the now Hermitian $i\phi^{3}$ theory
into a non-Hermitian theory with positive kinetic term. \ Though the form
obtained looks like the traditional one in the literature, a crucial
difference exists as our form is not $\mathcal{PT}$-symmetric and thus the the
$C$-operator regime followed in the literature to cure the ghost states
problem is no longer working. Another reason that this regime is not working
is that we have shown that the integration $\int d^{d}x\nabla^{2}\phi
(x)\pi^{2}$ vanishes while it has been set to non-zero value in the
literature. 

The soliton solutions obtained in this work are non-perturbative since they
are singular at the limit \ $g\rightarrow0$. Accordingly, in quantizing the
theory around the classical solutions, the physical quantities receives
non-perturbative corrections and thus turn them more accurate than the
quantization around the trivial vacuum. Moreover, the solitons are
non-topological  and in order to account for the stability of the
solitons we conjecture that the mother (equivalent) Hermitian theory has a
conserved Noether current. In fact, the current algorithms in the literature can not lead to
such kind of expectations and the theory deserves more careful analysis.

We think that the novel trend followed in this work will lead to a progress in
building up a concrete formulation of a strongly interacting scalar Higgs
mechanism.

\begin{acknowledgments}
We are very grateful to Hugh F. Jones for notifying us for the existence of some errors in the previous version of this work which we corrected  in the current version.
\end{acknowledgments}

\newpage 

\end{document}